%% file: CFTOT.tex
\documentclass[a4paper,11pt]{article}

\usepackage{jinstpub} 

\usepackage{xspace}
\usepackage{caption}
\usepackage{subfigure}
\usepackage{xcolor}
\usepackage{color}
\usepackage[normalem]{ulem}
\usepackage{bm}

\usepackage{lineno}

\title{\boldmath Optimization and first electronic implementation of the Constant-Fraction Time-Over-Threshold pulse shape discrimination method}

\author[a,b,1]{A. Roy,\note{Corresponding author.}}
\author[b]{D. Vartsky,}
\author[c]{I. Mor,}
\author[d]{C. Boiano,}
\author[d]{S. Brambilla,}
\author[d]{S. Riboldi,}
\author[a,e]{E. O. Cohen,}
\author[a,e]{Y. Yehuda-Zada,} 
\author[e]{A. Beck} 
\author[a]{and L. Arazi.}

\affiliation[a]{Unit of Nuclear Engineering, Ben-Gurion University of the Negev, Beer-Sheva, Israel}
\affiliation[b]{Department of Particle Physics and Astrophysics, Weizmann Institute of Science, Rehovot, Israel}
\affiliation[c]{Soreq Nuclear Research Center, Yavne, Israel}
\affiliation[d]{I.N.F.N. - Sezione di Milano, Italy}
\affiliation[e]{Nuclear Research Centre Negev, Beer-Sheva, Israel}

\emailAdd{arindam@post.bgu.ac.il}

\abstract{\input{src/abstract}}


\arxivnumber{1234.56789} 

\begin{document}
\maketitle
\flushbottom

\input{src/introduction}
\input{src/setup}

\input{src/classification_cftot}

\input{src/classification_cc}

\input{src/electronics}
\input{src/summary}

\appendix

\acknowledgments
\input{src/acknowledgement}

\input{src/bibliography}


\end{document}

%% file: src/abstract.tex
In this contribution we report on further investigations of the recently-evaluated Constant-Fraction Time-over-Threshold (CF-ToT) method for neutron/gamma-ray pulse shape discrimination (PSD). The superiority of the CF-ToT PSD method over the constant-threshold (CT-ToT) method was previously demonstrated, down to low neutron energy thresholds of 100 keVee. Here, we report on a quantitative comparison between the traditionally used Charge Comparison (CC) method and the CF-ToT method using a stilbene scintillator coupled to a silicon photomultiplier, implementing an offline analysis of recorded fast-neutron and gamma-ray waveforms. An optimization of the constant fraction value indicates that a 20$\%$-fraction yields the optimum figure-of-merit (FOM) and gamma-ray peak-to-valley (P/V) ratio. The results obtained for a particle energy threshold of 100 keVee show that the FOM and P/V values achieved with the CF-ToT method are superior to those obtained using the standard CC method. In addition, a first electronic implementation of the CF-ToT method was performed using simple circuitry suitable for multichannel architecture. Initial results obtained with this circuit prototype are presented.

%% file: src/introduction.tex
\section{Introduction}
\label{sec:intro}

Time-Over-Threshold (ToT) is defined as the time interval during which a detected pulse exceeds a specific voltage threshold. The measurement of ToT is fast and simple to implement using a voltage comparator, and in many circumstances can effectively replace a measurement of pulse height \cite{kipnis}. ToT depends on both the amplitude and pulse shape in a non-linear fashion. Therefore, several methods such as dynamic time-over-threshold (DToT)~\cite{yonggang,orita}, time-over-linear-threshold (TOLT)~\cite{song} and multiple thresholds (MToT)~\cite{amiri,georgakopoulou} have been proposed for obtaining a linear ToT-charge relationship, with considerable work performed on the use of ToT for neutron/gamma-ray pulse shape discrimination (PSD) in organic liquid and solid scintillators~\cite{amiri,jastaniah}. In applications such as fast-neutron multiplicity counting (FNMC)~\cite{fulvio}, large-area neutron cameras for fast-neutron resonance radiography (FNRR)~\cite{dangendorf,mor,vartsky,noam}, fast-neutron computed tomography (CT)~\cite{adams} and Gamma-Resonance Absorption (GRA) radiography~\cite{vartsky2}, multiple detector channels (hundreds to thousands) accompanied by front-end electronics that can provide good timing and neutron/gamma discrimination without digitization of the full waveform are essential. Existing multi-channel  Application Specific Integrated Circuits (ASICs) are all based on the CT-ToT~\cite{abbaneo,rolo} method. 

In a recent publication~\cite{roy} we analysed a different variant of ToT, using a constant fraction method (CF-ToT), demonstrating its superior performance in neutron/gamma PSD compared to CT-ToT for energies in the range 100-1000 keVee.
Since CF-ToT is independent of pulse amplitude and varies solely with the pulse shape, it can potentially serve as a better alternative for the design of future multi-channel PSD circuits that avoid pulse digitization.

In this work we present a comprehensive comparison between the conventional, widely used, Charge Comparison (CC) PSD method---which requires digitization of the full waveform---and the CF-ToT method, using a stilbene scintillator coupled to a silicon photomultiplier (SiPM) light sensor with offline analysis of recorded fast-neutron and gamma-ray waveforms. Additionally, we present preliminary results obtained with a first prototype of an electronic circuit that performs real-time CF-ToT PSD. 

%% file: src/setup.tex
\section{Experimental Setup}
\label{sec:setup}

\begin{figure}
    \centering
    \includegraphics[scale=0.52]{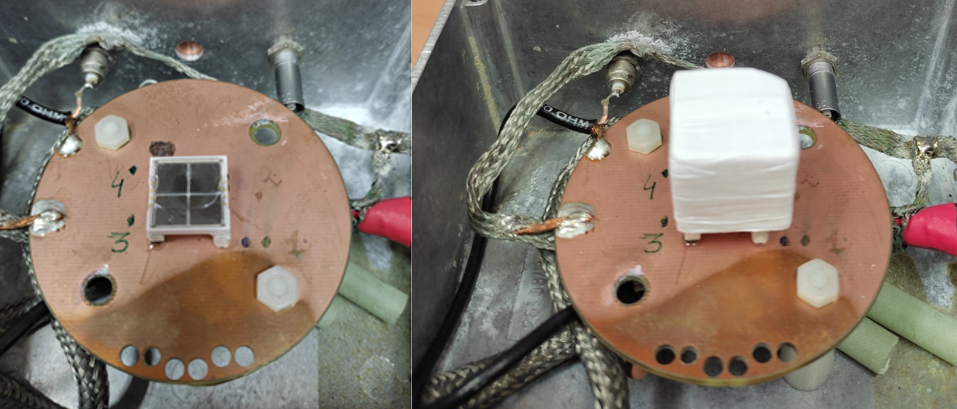}
    \caption{20$\times$20$\times$20 mm$^{3}$ stilbene crystal coupled to a 12$\times$12 mm$^{2}$ quad SiPM sensor, manufactured by Hamamatsu.}
    \label{fig:i}
\end{figure}

\begin{figure}
   \centering
   \includegraphics[scale=0.57]{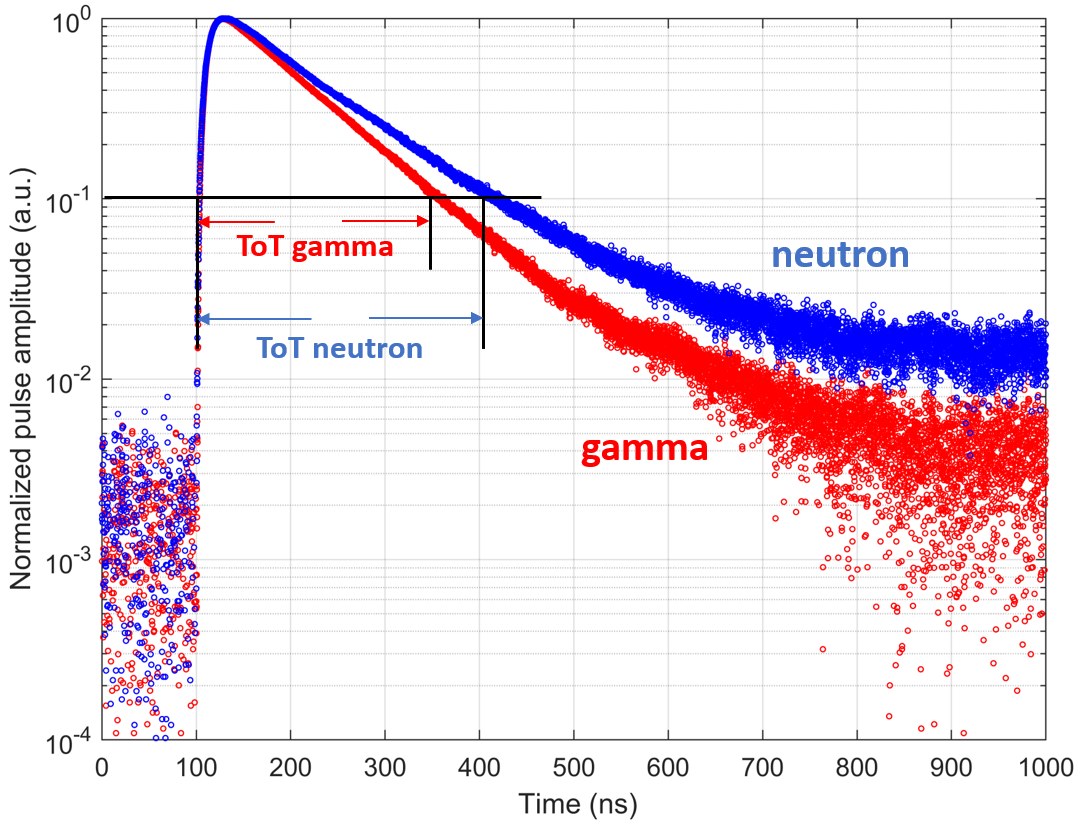}
   \caption{\label{fig:ii} An illustration of the time-over-threshold for a pair of normalized neutron and gamma-ray pulses obtained from the stilbene crystal, coupled to a S13371-6050CQ-02 Hamamatsu Quad VUV4 SiPM.}
\end{figure}

The stilbene crystal used for the experiment was a 20$\times$20$\times$20 mm$^{3}$ Scintinel detector, produced by Inradoptics. The crystal was wrapped with Teflon tape on all sides, except for the one coupled to a quad-SiPM light sensor (Hamamatsu Quad VUV4 MPPC, model S13371-6050CQ-02 ~\cite{ohashi}. The quad SiPM is comprised of four SiPM segments, each with an area of 6$\times$6 mm$^{2}$, with a 0.5 mm gap between segments; it has 13,923 pixels per segment and a geometrical fill-factor of $\sim$ 60$\%$. The window is made of quartz and the pixel pitch is 50~$\mu$m~\cite{rolo}. The SiPM operating voltage was maintained at -57 V. Figure~\ref{fig:i} (left) shows the quad SiPM and Figure~\ref{fig:i} (right) -- the stilbene crystal wrapped with Teflon tape and coupled to the SiPM array. The detector signals were digitized using a Lecroy Waverunner 610ZI oscilloscope with a sampling rate of 20 GS/s. Evaluation of the performance of the PSD methods was done offline. The pulses from each SiPM segment were summed to yield the total pulse value. The system was energy-calibrated with $^{137}$Cs and $^{60}$Co gamma-ray sources, as well as with the 4.43 MeV gamma ray from an AmBe source. All results are expressed in electron-equivalent energy (keVee). Neutrons were produced by a 100~mCi AmBe source, shielded with a 5~mm-thick lead enclosure to suppress the intense 59.54 keV gamma ray. 10,000 pulses were accumulated and stored for further analysis. A typical neutron and gamma-ray pulse obtained with the stilbene-SiPM configuration is shown in Figure~\ref{fig:ii}. The pulses have been normalized to peak amplitude in order to highlight the difference in the trailing edge of the pulse. The time over threshold (ToT) definition for the two pulses for a constant peak fraction of 10$\%$ of the peak amplitude is also illustrated.

%% file: src/classification_cftot.tex
\section{Pulse analysis by the CF-ToT method}
\label{sec:cftot}

Offline analysis of the 10000 digitized pulses by the CF-ToT method was performed by the following procedure:
\begin{itemize}
\item Baseline correction.
\item Finding the peak position and the peak amplitude $A$.
\item Applying a selected fraction $f$ on $A$.
\item Determining the ToT value defined as the width of the pulse above the voltage threshold $f\cdot A$.
\end{itemize}

\begin{figure}
   \centering
   \includegraphics[width=15cm]{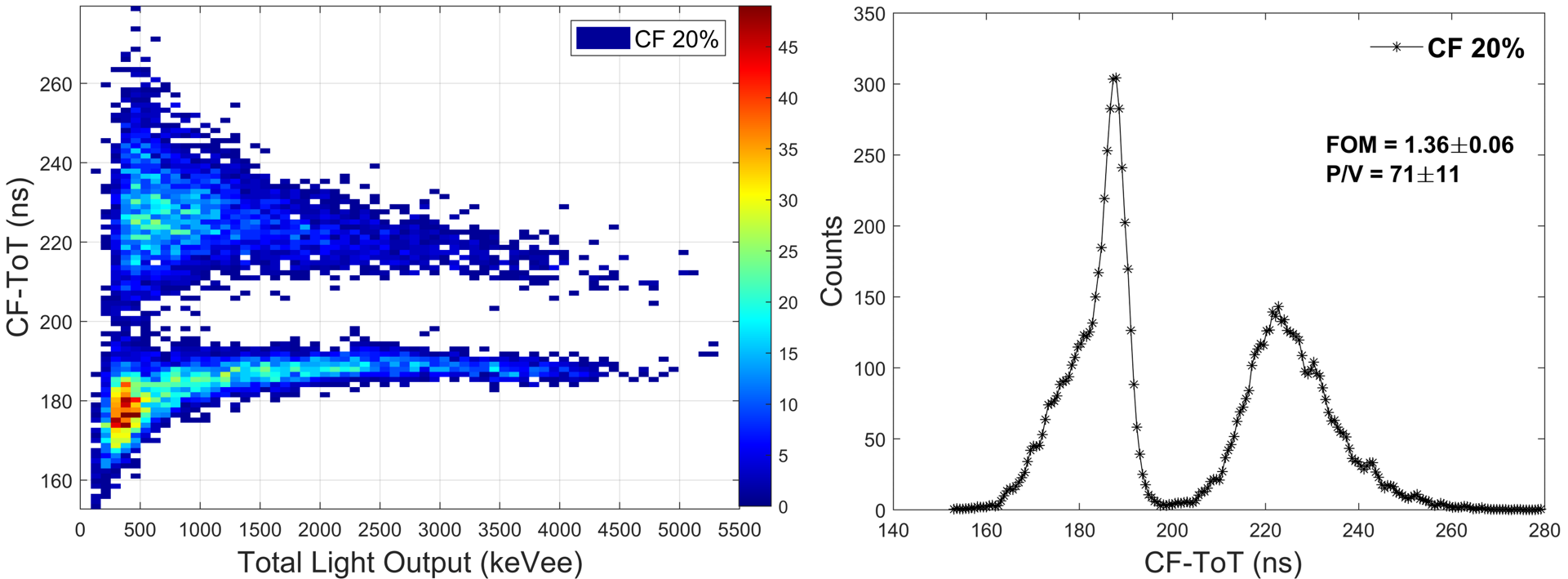}
   \caption{\label{fig:iii} Left: CF-ToT as a function of the total light output expressed in keVee; right: frequency distribution of the CF-ToT values. The selected fraction was 20$\%$ and the energy threshold was 100~keVee.}
\end{figure}

This procedure was applied to all pulses above an energy threshold of 100 keVee. The CF-ToT value \textemdash the duration of the pulse above the $f\cdot A$ threshold, expressed in nanoseconds\textemdash was utilized as the PSD parameter. Figure~\ref{fig:iii} (left) shows the CF-ToT as a function of the total light output (event energy), expressed in keVee for a constant fraction of 20$\%$. The lower band comprises gamma-ray events and the upper one \textemdash neutrons. The event frequency distribution as a function of CF-ToT is shown in Figure~\ref{fig:iii} (right). This curve represents a projection of the entire 2D histogram shown on the left on the CF-ToT axis, and the width of the bands is dominated by low-energy events.

The effectiveness of any PSD parameter can be quantified by a figure-of-merit (FOM), defined as the ratio of the distance between the neutron and gamma-ray peak positions in the CF-ToT frequency distribution, Figure~\ref{fig:iii} (right), and the sum of the full widths at half maxima of the two peaks. Another measure of PSD quality is the peak-to-valley ratio (P/V), defined here as the ratio of the maximum value of the gamma-ray peak to the average of the four lowest points of the valley of the distribution. The FOM and P/V values obtained for the 20$\%$ CF-ToT fraction are indicated on the frequency distribution plot in Figure~\ref{fig:iii} (right). Although the P/V ratio depends on the relative intensity of the gamma rays and neutrons measured by the detector and is therefore system- and source-dependent, the use of this parameter for comparison of PSD methods applied on the same detector-source configuration seems a valid approach.

\begin{figure}
   \centering
   \includegraphics[width=15.3cm]{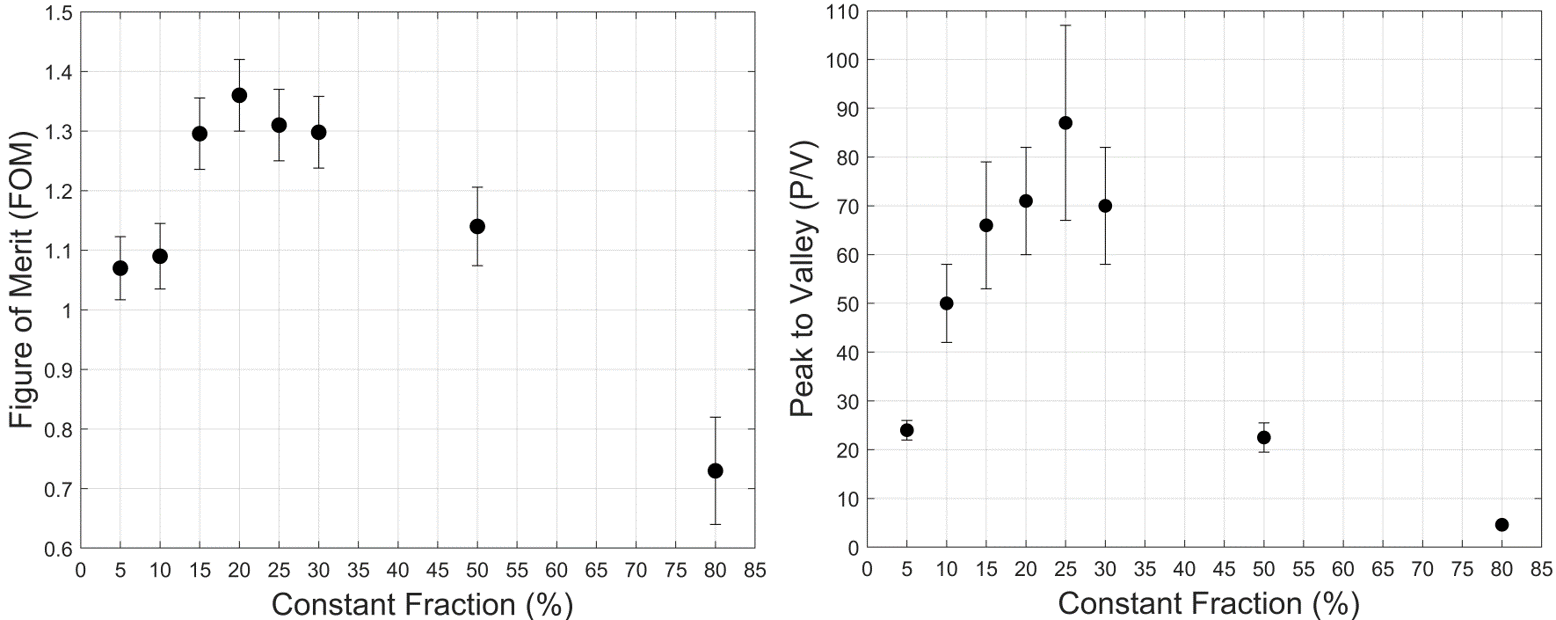}
   \caption{\label{fig:iv} FOM (left) and P/V ratio (right) for the CF-ToT method as a function of the applied constant fraction. The energy threshold was 100~keVee.}
\end{figure}

Figure~\ref{fig:iv} shows the dependence of the FOM (left) and P/V (right) on the applied constant fraction value.  The optimal constant fraction is in the range of 20-25$\%$, with FOM = 1.37$\pm$0.06 and P/V = 87$\pm$20. The large errors in P/V result from the low statistics in the valley.

%% file: src/classification_cc.tex
\section{Pulse analysis by the CC method}
\label{sec:cc}

The same set of 10,000 digitized pulses were analysed based on the standard CC method. The ratio of the charge in the tail of the pulse, {\it Q$_{tail}$}, to the total charge {\it Q$_{total}$} is defined as the CC$_{PSD}$ parameter

\begin{equation}
\label{eq:y:1}
CC_{PSD} = Q_{tail}/Q_{total} = (Q_{long}-Q_{short})/Q_{long}
\end{equation}

{\noindent}where {\it Q$_{short}$} and {\it Q$_{long}$} are the integrated charge values evaluated over the short ({\it t$_{short}$}) and long ({\it t$_{long}$}) time windows respectively. The time windows were selected to optimize the FOM and P/V ratio as defined above. {\it t$_{short}$} and {\it t$_{long}$} values of 183 ns and 1.2 $\mu$s were used in this work, including a pre-gate of 100 ns. 

\begin{figure}[!ht]
   \centering
   \includegraphics[width=15.3cm]{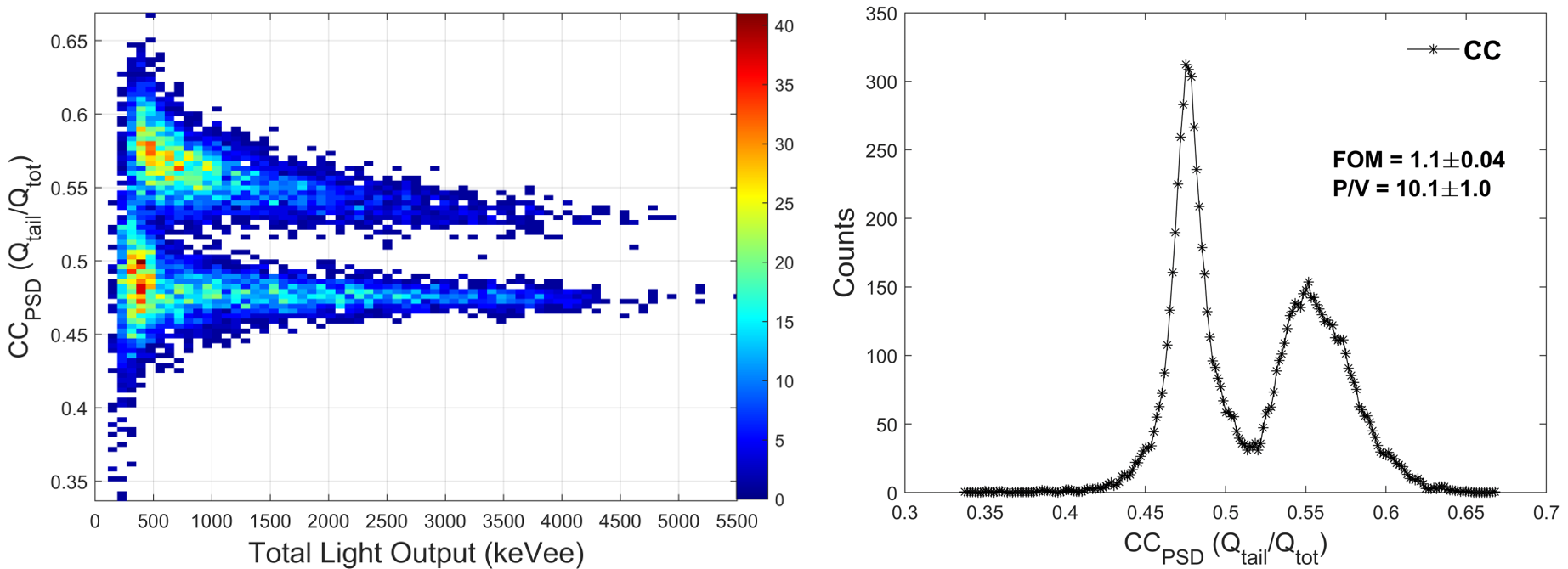}
   \caption{\label{fig:v} Left: CC$_{PSD}$ as a function of the total light output (event energy), expressed in keVee; right: frequency distribution of CC$_{PSD}$. The energy threshold was 100 keVee.}
\end{figure}

Figure~\ref{fig:v} (left) shows the PSD parameter evaluated for the CC method as a function of the total light output (event energy), expressed in keVee and Figure~\ref{fig:v} (right) shows the frequency distribution. The FOM value obtained with the CC method was 1.1$\pm$0.04 and the P/V ratio was 10.1$\pm$1.0, significantly lower than those obtained by the CF-ToT method.

%% file: src/electronics.tex
\section{Electronic implementation of the CF-ToT method}
\label{sec:electronics}

\begin{figure}[!ht]
   \centering
   \includegraphics[scale=0.38]{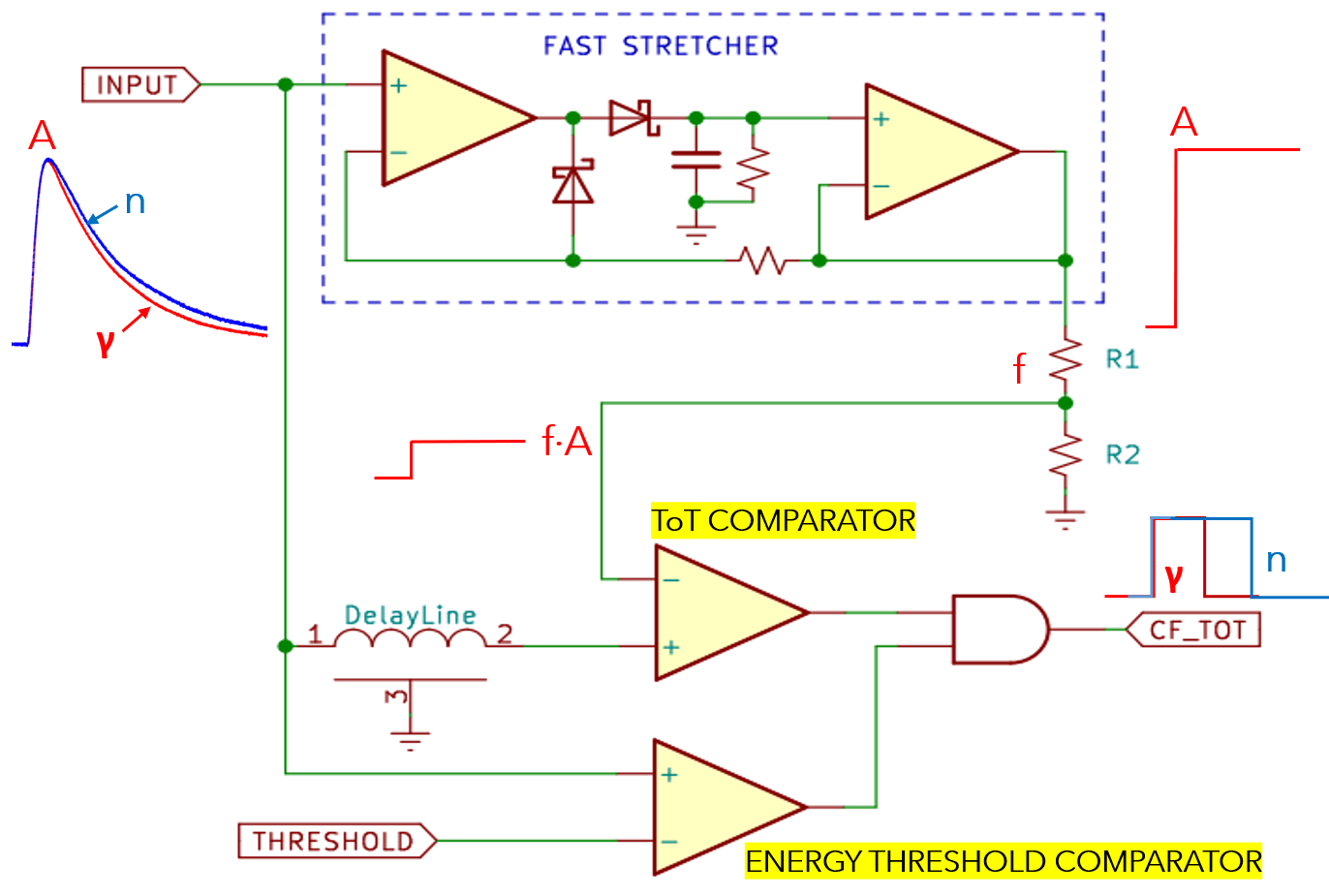}
   \caption{\label{fig:vi} Block diagram of the CF-TOT method implemented in a circuit, along with a pictorial representation of the waveforms at every stage.}
\end{figure}

To the best of our knowledge, an electronic circuit that performs constant-fraction triggering on the trailing-edge of a pulse has not yet been implemented. The block diagram illustrated in Figure~\ref{fig:vi} describes schematically a method for the implementation of the CF-ToT method in an analog circuit.

{\noindent} The working principle of the circuit is as follows:
\begin{itemize}
    \item The detector signal is split into three branches.
    \item The first branch is delivered to a lower level energy threshold comparator, whose output enables the CF-ToT output. Events with an amplitude below the threshold set by the comparator will not be processed.
    \item The second branch is delivered to the (+) input of a ToT comparator through a suitable delay.
    \item The third branch is delivered to a fast stretcher, whose function is to provide an output pulse of the same amplitude as the peak value of the input pulse $A$, but stretched in duration. 
    \item The resulting stretched signal $A$ is then attenuated by a resistive voltage divider to the desired fraction $f\cdot A$ and is delivered to the (-) input of the ToT comparator.
    \item The input signal and the attenuated stretched signal are compared in the TOT comparator to produce a logic signal of the duration of the time-over-threshold (ToT). 
    \item Since every signal is always compared to a constant fraction of its own peak amplitude, the ToT is independent of pulse amplitude and depends only on the pulse shape. 
\end{itemize}

\begin{figure}[!ht]
   \centering
   \includegraphics[scale=0.55]{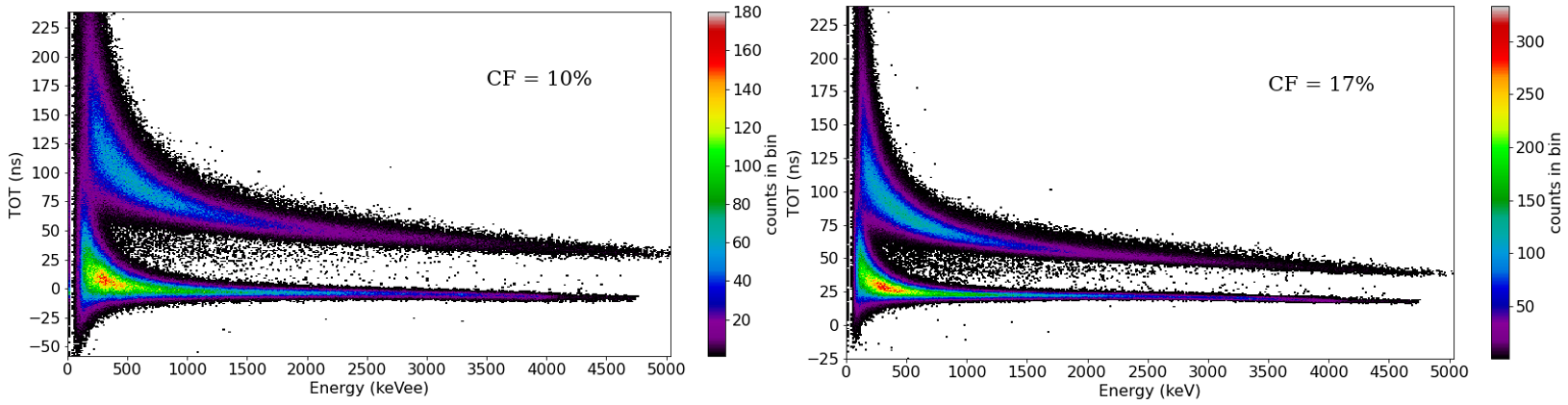}
   \caption{\label{fig:vii} CF-ToT as a function of the energy expressed in keVee, obtained using the circuit shown in Figure \ref{fig:vi} for constant fractions of 10\% (left) and 17\% (right), at an energy threshold of 100 keVee.}
   
\end{figure}

\begin{figure}[!ht]
   \centering
   \includegraphics[width=15cm]{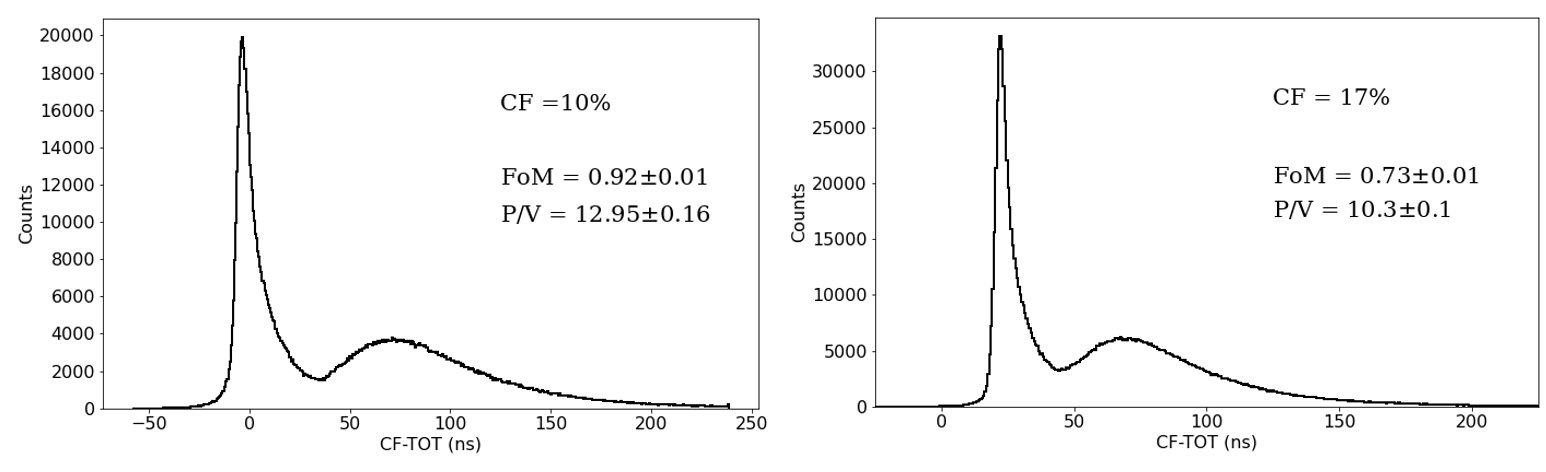}
   \caption{\label{fig:viii} CF-ToT frequency distributions corresponding to Figure \ref{fig:vii}, at 100 keVee energy threshold.}
\end{figure}

A preliminary circuit has been designed by the INFN Milano group, based on their design of a fast peak stretcher~\cite{boiano}. Real-time CF-ToT pulse shape discrimination measurements were performed with this circuit. An AmBe source of 10 mCi strength, shielded with a 10 mm thick layer of lead and positioned at a distance of 100 mm from the detector was used as the neutron source. Gamma rays from a $^{22}$Na source were used for the energy calibration in keVee. In this study, the stilbene detector described in Section~\ref{sec:setup}, was placed at the center of a SensL J60035-64 SiPM array (8$\times$8 array of 6$\times$6 mm$^{2}$). The signal was a sum of the 64 SiPMs, using a summing card described in ~\cite{boiano2}. The use of this large array to observe light from a 20$\times$20 mm$^{2}$ stilbene detector is sub-optimal since the detector covers only about 10 SiPM segments. The rest of the SiPMs do not see the light signal but still contribute to the noise. (This will be improved in future studies with an optimized summing card.) 
In order to analyse the circuit performance both the CF-ToT logic signal and the detector pulse were digitized by the CAEN V1730 digitizer to provide the ToT and the energy signal. Figure~\ref{fig:vii} shows the first experimental results of CF-ToT obtained with a real-time circuit as a function of the energy, for two different constant fractions, 10\% shown on the left and 17\% shown on the right. Figure~\ref{fig:viii} shows the respective CF-ToT frequency distributions for the 10\% (left) and 17\% (right) fractions. 
The FOM values obtained with an energy threshold of 100 keVee were 0.92$\pm$0.01 and 0.73$\pm$0.01 and the P/V ratios were 12.95$\pm$0.16 and 10.3$\pm$0.1 for the 10\% and 17\% fractions, respectively. The initial results are promising and circuit optimization is currently ongoing.

%% file: src/summary.tex
\section{Summary and discussion}
\label{sec:summary}

In our previous work, we compared the PSD performance of the CF-ToT method with the CT-ToT method and demonstrated that the former is superior and can be used down to 100 keVee particle energy thresholds, while the latter could only be used above relatively high thresholds of 1000 keVee. 
In this work, we compare the PSD performance of the traditionally used CC method with the CF-ToT method in terms of the FOM and P/V values obtained. The offline analysis of 10000 recorded waveforms indicated that the optimal CF-ToT fraction for our configuration is $\sim20-25\%$. The digitized neutron and gamma pulses illustrated in Figure~\ref{fig:ii} offer an explanation for the existence of an optimal CF-ToT fraction. At high fractions (30-80$\%$), one samples mainly the amplitude contribution of the dominant fast decay component, which is similar for neutrons and gamma rays. In the case of very low fractions, the noise becomes dominant and the fact that the pulse becomes nearly horizontal broadens the triggering position. It appears that for our detector configuration, PSD obtained with the CF-ToT method is significantly better than that obtained by the CC method as evident from the higher values of the associated figure-of-merit and peak-to-valley ratio.

It is important to determine how each method classifies the events into neutrons and gamma rays. A threshold cut is used at the minima of the event frequency distributions of both CF-ToT [Figure~\ref{fig:iii} (right)] and CC$_{PSD}$ [Figure~\ref{fig:v} (right)]. All events below the threshold are classified as gamma rays and the events above it as neutrons. 97\% of the events were classified identically by the two methods. The remaining 3\%, defined as neutrons by the CC classification method are interpreted as gamma-rays by the CF-ToT method. Note that in order to classify the neutron/gamma-ray events with high confidence, it is preferable to use the time-of-flight method.

The FOM values appearing in literature, achieved with the CC method for the stilbene-SiPM configuration vary greatly. FOM values of 1.15~\cite{ruch}, 1.17~\cite{steinberger}, 1.37 and 2.13~\cite{pozzi}, 1.6, 1.76~\cite{chergui} and 1.5~\cite{ferrulli} and 1.27~\cite{boo} have been reported. This can be attributed to variations in the SiPM sensor types and the differences in experimental conditions, such as detector size, light collection efficiency, energy thresholds and energy cuts.
The optimal FOM value for CC obtained using our detector configuration is 1.1$\pm$0.04, which is on the lower side of most of the reported CC FOM values. We worked exclusively with a 100 keVee, low-energy threshold, for both the CF-ToT and the CC methods. Even for mono-energetic neutrons, the resulting proton energies extend down to zero energies. Therefore, one should try to operate with the lowest possible energy threshold and in our opinion, FOM results at higher energy thresholds or energy cuts present an incomplete picture. Since we use the same set of digitized pulses and the same energy threshold for neutron/gamma-ray discrimination by the two methods, our comparison of the CF-ToT and the CC method seems fair.

Beyond comparing the performance of the CF-ToT and CC methods, a primary aim of this study was to develop a simple electronic circuitry that would permit obtaining PSD without pulse digitization. In this work we introduced a simple solution for electronic implementation of CF-ToT. The circuit developed here already shows that reasonable separation of neutrons from gamma rays can be obtained with a threshold of 100 keVee. Work on optimizing the circuit is ongoing.

%% file: src/acknowledgement.tex
The authors would like to thank Dr. Shikma Bressler from the Department of Particle Physics $\&$ Astrophysics of the Weizmann Institute of Sciences for lending us the stilbene scintillator.  The work was performed under grant no. 3-16313 from the Israel Ministry of Science and Technology.